# LSCHC: Layered Static Context Header Compression for LPWANs

Khaled Q. Abdelfadeel, Victor Cionca, Dirk Pesch
Nimbus Center, Cork Institute of Technology, Ireland
khaled.abdelfadeel@mycit.ie, victor.cionca@cit.ie, dirk.pesch@cit.ie

## ABSTRACT

Supporting IPv6/UDP/CoAP protocols over Low Power Wide Area Networks (LPWANs) can bring open networking, interconnection, and cooperation to this new type of Internet of Things networks. However, accommodating these protocols over these very low bandwidth networks requires efficient header compression schemes to meet the limited frame size of these networks, where only one or two octets are available to transmit all headers. Recently, the Internet Engineering Task Force (IETF) LPWAN working group drafted the Static Context Header Compression (SCHC), a new header compression scheme for LPWANs, which can provide a good compression factor without complex synchronization. In this paper, we present an implementation and evaluation of SCHC. We compare SCHC with IPHC, which also targets constrained networks. Additionally, we propose an enhancement of SCHC, Layered SCHC (LSCHC). LSCHC is a layered context that reduces memory consumption and processing complexity, and adds flexibility when compressing packets. Finally, we perform calculations to show the impact of SCHC/LSCHC on an example LPWAN technology, e.g. LoRaWAN, from the point of view of transmission time and reliability.

## 1. INTRODUCTION

Low Power Wide Area Networks (LPWANs) [1] [2] such as LoRa [3], Sigfox [4], and NB-IoT [5], with their ability to connect billions of low-cost devices are considered a key driver for the Internet of Things (IoT). LPWANs enable cost-effective data collection for many IoT applications, from agriculture to smart cities. Data can be transmitted over many kilometers, which simplifies network topologies, protocols, and deployment, however, these technologies still function as islands of connectivity, independent of each other [6]. These islands can all be connected into one IoT by introducing the native IoT protocol stack i.e. Internet Protocol (IP), User Datagram Protocol (UDP), and Constrained Application Protocol (CoAP) into these technologies.

The challenge of supporting this native IoT stack over LPWANs is the limited Layer 2 characteristics [6] such as small frame sizes in the order of tens of octets, very low duty cycle, low data rate, and with mostly upstream traffic favored over downstream transmission to allow devices to spend most of their time in deep sleep mode. Due to these characteristics, LPWANs should be considered as a separate class of wireless networks in order to find new approaches/solutions to support the IoT protocol stack. Therefore, the Internet Engineering Task Force (IETF) formed the LPWAN working group in 2016 to address the adaptation of IPv6, UDP, and CoAP over LPWANs [7].

One of the main problems in adapting the IoT stack for LPWANs is the large protocol header overhead, e.g. 40 octets for IPv6 header, 8 octets for UDP header, and 4 octets for the short fixed-length CoAP header, compared to the very limited frame size of the LPWANs. In some cases, one or two octets would be available to transmit all these headers. Therefore, an efficient header compression scheme is crucial in order to adapt the IoT stack to LPWANs, taking advantage of the unique characteristics of LPWANs such as the star-topology and that the data flows are known-in advance due to pre-programmed applications.

The IETF LPWAN working group proposes Static Context Header Compression (SCHC), a new header compression scheme for the IPv6/UDP headers [8], and for the CoAP header [9]. SCHC is based on a shared static context that does not change with time, thus, avoiding complex synchronization, which is the most extensive operation in header compression. SCHC can compress the headers of IPv6, UDP, and CoAP down to a few bits by omitting known and redundant information, thus, reducing the network overhead and speeding up the transmission of packets, which leads to a lower power consumption. However, the drafts in [8] and [9] do not provide any evaluation of the SCHC scheme in terms of compression factor and transmission time.

The contributions of this paper can be summarized as follows: Firstly, we provide a detailed overview of the SCHC mechanism. As far as we know, this work is the first paper that investigates the SCHC mechanism. Secondly, we present an enhancement of SCHC, the layered SCHC (LSCHC) scheme. LSCHC is a layered context that saves memory in constrained devices, reduces the processing complexity and adds flexibility when compressing packets. Thirdly, we provide some insights from our implementation experiences of the SCHC and LSCHC schemes. Fourthly, we present a performance evaluation of the SCHC/LSCHC. The SCHC/LSCHC scheme is compared with the IPHC scheme [11], another IETF header compression for constrained networks. Finally, we show the impact of SCHC/LSCHC on LoRa technology, as an example of LPWANs, from the point of transmission time and reliability.

The remainder of this paper is organized as follows: Section II highlights some of the related work in the area of header compression. Section III presents an overview of the SCHC mechanism and introduces the layered SCHC. We describe our implementation experiences in Section IV. Section V explains the evaluation process and discusses the results. Finally, Section VI presents our conclusion and our future work.

## 2. Related Work

Generally speaking, header compression schemes can be divided into three categories: stateless, stateful or hybrid. The stateless schemes are simple encoding rules, where the encoding does not depend on a flow, default values are assumed instead. This leads to stateless schemes not achieving a good compression factor with multiple flows. Whereas, stateful schemes build context(s) for each flow to be compressed. The contexts require memory, result in processing time overhead and consume additional bandwidth for building and synchronization. However, they can achieve good compression factors with multiple flows simultaneously. The condition of the communication medium is critical for the stateful schemes where a lossy medium can cause a desynchronization between the compression and the decompression sides. This leads to additional delay in building and recovering contexts, thus, adding delay to compress/decompress the packets. In hybrid schemes, the stateless and the stateful methods are combined into one scheme. By default it is stateless. In case the compression factor is low, it switches to stateful compression.

RFC 4944 [10] defined the 6LOWPAN_HC1 and the 6LOWPAN_HC2 schemes as stateless header compressions for the IPv6 header and the next header, respectively over IEEE 802.15.4. HC1 assumes default values for the IP version, traffic class, and flow label fields. The next header segment can be compressed down to two bits and the hop limit value is carried in-line. HC1 assumes the payload length can be inferred from the header of the lower layer and the source and the destination addresses are unicast link-local addresses, where the Interface Identifier (IID) is directly derived from the IEEE 802.15.4 addresses.

HC1 is an effective scheme for unicast link-local communications but has a very limited effect on global and multicast addresses. Therefore, HC1 is commonly used for local protocol interactions such as IPv6 neighbor discovery or routing protocols. In the best case, HC1 can compress the IPv6 header down to two octets (one octet for the HC1 encoding and one octet for the hop limit) in the case of unicast link-local communication. However, when the destination address is a multicast address or a global address, the HC1 requires the full 128-bit address to be carried in-line.

HC1 extends support for compressing the next header by using HC2, but only for UDP, TCP, and ICMPv6. However, [10] describes only how the UDP header can be compressed, where the UDP length can be inferred from the header of the lower layer. The commonly used port numbers in the range from F0B0 to F0BF can be compressed down to four bits. The UDP checksum is carried in-line. HC2 assumes the IPv6 and UDP headers are contiguous headers. Therefore, HC2 cannot compress the UDP header in case an IPv6 extension header is present. In the best case, HC2 can compress the UDP header down to four octets.

6LOWPAN HC1 and HC2 are insufficient for most practical uses of IPv6 in 6LoWPAN. As a result, RFC 6282 [11] defined the encoding formats, LOWPAN_IPHC and LOWPAN_NHC, for compressing the IPv6 header and the next header, respectively to overcome the shortcomings of the above schemes. IPHC is an example of a hybrid header compression scheme that employs the stateless compression with the link-local address using 13 bits encoding and employs stateful compression, when necessary, to compress the global and multicast addresses using an additional 8 bits to store shared contexts for arbitrary prefixes. The context of IPHC allows up to sixteen network prefixes to be compressed when communicating with external networks, however, [11] does not specify any way to build or maintain this context. In the best case, the IPHC can compress the IPv6 header down to two octets (dispatch and encoding) with link-local communication and three octets (two octets for dispatch and encoding, and one octet for stateful context) for the multicast and global communications.

IPHC supports compression of the next header using NHC. NHC can compress any arbitrary next header, however, the standard in [11] covers UDP and some of the IPv6 extension header only. NHC assumes the UDP length can be inferred from the lower layer and the checksum can be recalculated at the decompressor. In the best case, the NHC can compress the UDP header down to one octet. However, for most practical cases, it compresses the UDP header down to five octets (one octet for encoding and four octets for the ports).

RObust Header Compression (ROHC) [12] is a generic/versatile header compression scheme that can work on different headers such as IP, UDP, TCP, and RTP. RoHC is a stateful scheme in which the compressor and the decompressor share a context. To build the context, ROHC assumes the packets are classified firstly into flows before being compressed, thus, the RoHC takes advantage of the information redundancy for packets belonging to the same flow, where the static redundant information such as source address and destination address etc. are transmitted in the first packet only. Variable information such as identifiers and sequence numbers etc. are sent in a compressed form to save bandwidth. Once a packet is classified as belonging to a flow, the compression is performed according to a profile. A profile defines the compressing function for the different fields in the network header(s). There are several compression profiles for different flows, for example, IP only, IPv6/UDP, IP/UDP/RTP and IP/TCP etc.

To ensure context synchronization, ROHC has three modes of operation: Unidirectional mode (U-mode), bidirectional Optimistic mode (O-mode), and bidirectional Reliable mode (R-mode). The U-mode specifies compression over a unidirectional link in which the packets are sent from the compressor to the decompressor. In order to handle the potential errors, the compressor sends periodic updates of the flow context to the decompressor. The O-mode is similar to the U-mode, however, over a bidirectional link. The O-mode uses a feedback channel to send optional recovery requests and acknowledgments of significant context updates from the decompressor to the compressor. Whereas R-mode uses extensively the feedback channel and depends on a strict logic that ensures loss-free context synchronization. ROHC can compress header(s) of different flows very effectively. In the best case, the flow header(s) can be compressed down to one or two octets.

The majority of LPWANs are highly constrained networks with very limited frame sizes, in some cases, only one or two octets are available to transmit all the headers. The aforementioned schemes such as HC1/HC2 or IPHC/NHC cannot achieve the required level of compression for LPWANs. Furthermore, they do not consider the application layer header e.g. CoAP. Therefore, a new header compression scheme is needed that must be able to compress the application layer header such as CoAP besides compressing the lower layer headers. The new scheme should exploit the unique specifications of LPWAN such as the star-topology and the fact that flows are known in advance due to pre-programmed applications. Although the ROHC may provide the required level of compression, the learning and the synchronization introduce communication overheads that are prohibitive for LPWANs. Furthermore, RoHC entails a significant amount of implementation complexity, which translates directly to an increased amount of processor and memory utilization on devices.

## 3. Static Context Header Compression

The IETF LPWAN working group was relatively recently formed to study the adoption of Internet technologies to LPWANs [7]. It proposed the SCHC as a header compression scheme for LPWAN, covering so far, the IPv6/UDP headers [8], as well as the CoAP header [9]. However, SCHC is not limited to LPWANs and could be extended to other protocol stacks. The work is still ongoing, but it has now reached a good level of maturity.

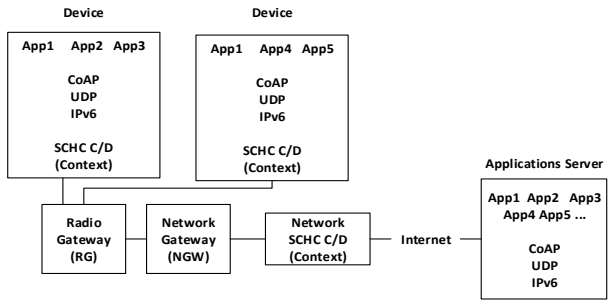

Figure 1. SCHC architecture

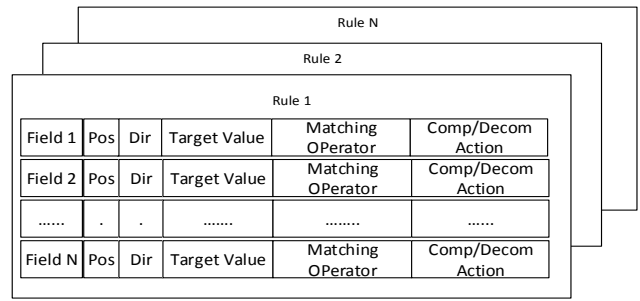

Figure 2. SCHC compression /decompression context

The SCHC is a stateful header compression that is based on building a shared static context among the network elements to compress/decompress the header(s). *Static* means that the content of the context does not change with time, avoiding complex synchronization, which is the most resource consuming operation in the other stateful header compression schemes such as RoHC. SCHC is based on the fact that the traffic flows are mostly known in advance since devices embed fixed built-in applications. Therefore, a shared static context may be stored on devices within the LPWAN to compress/decompress the previously known flows.

Figure 1 shows the architecture of a typical LPWANs with devices running applications that produce/accept different flows. These flows are compressed/decompressed using the device SCHC Compression/Decompression (C/D) unit and the network SCHC C/D unit. The architecture shows the bidirectional communication between the devices and the application server where data travels through the Radio Gateway (RG) and the Network GateWay (NGW). The one-hop links between end devices and the RG are constrained, however, the links between the RG and the application server, through the NGW, are not constrained e.g. IP Ethernet link. The network SCHC C/D can be part of the NGW or outside if a tunnel between the NGW and the SCHC C/D is established.

The context, as shown in Figure 2, consists of rules that are lists of fields. A rule targets a header or more to be compressed, e.g., a rule for the IPv6 header, IPv6/UDP/CoAP headers, or IPv6/ICMP headers and so on. Each rule is defined using a *rule ID*. The rule ID is sent to the other end followed by the information resulting from the compression of the header fields. A field corresponds to a segment in the potential header to be compressed, describing the action used to compress/decompress this field. A field in a rule contains several entries: Field ID, which is a unique value to define the field; Field Position, which indicates which instance is targeted in case several instances of the field exist; Direction Indicator, which specifies the direction of the packet that could be upstream, downstream or bidirectional; Target Value, which is the value compared with the packet header value; Matching operator, which is the operator used to make the comparison between the target value and the packet header value; and C/D action, which describes the process of compression and decompression of this field.

The SCHC draft [8] defines a set of basic matching operators such as *equal*, which means the packet header value equals the target value, and *ignore*, which means no check is done between the packet header value and the target value. The C/D action has a different meaning at the C/D and depends on the used matching operator. The *not-sent* action is usually used with the *equal* matching operator. The compressor does not send the packet header value and the decompressor uses the value stored in the context. The drafts in [8] and [9] provide further information about the suitable matching operators and C/D actions for the IPv6/UDP headers and CoAP header, respectively.

An example of composing a rule to target a specific IPv6/UDP flow is shown in Figure 3. The example shows the different fields of each header and the target value, matching operator, C/D action that are used in each field. Compression is performed when a packet's header(s) matches one of the rules in the context. Then, the compressor uses the compression actions of the matched rule to compress the header(s) and sends the rule ID to the decompressor. Subsequently, the decompressor uses the sent rule ID to identify the used rule and applies the decompression actions to the received packet. The size of the rule ID can be set based on the number of flows and/or LPWAN technology. The used rules and their IDs must be identical at the device SCHC C/D and the network SCHC C/D. However, the draft [8] does not specify how this might happen.

The SCHC can compress several flows simultaneously with a great compression level based on the static shared context. The SCHC does not need complex synchronization, thus, it is suitable for constrained networks such as LPWANs. SCHC also is a generic solution that could be extended to other protocol stacks. In the best case, SCHC can compress the IPv6/UDP/CoAP headers down to a few bits, which equal the size of the rule ID.

## 3.1 Layered SCHC

SCHC uses a *single* static context, as shown in Figure 2, to save the different rules, and rules can cover several layers of the network stack. However, we argue that this is not the most efficient method to represent the rules and the method is likely to increase memory usage in the constrained devices. To explain the issue, assume we have two IPv6/UDP flows to the same IPv6 host, which means we have one IPv6 header but two different UDP ports. This can happen when the target host runs two concurrent applications on different UDP ports. SCHC would compose a rule for each flow, resulting in a context similar to that shown in Figure 4.

Although the fields of the IPv6 header are the same in rule one and rule two, SCHC stores two versions of the fields for the IPv6 header

| Field ID | Pos | Dir | Target Value | Matching Operator | C/D Function |
|---|---|---|---|---|---|
| IPv6 V | 0 | B | 6 | equal | not-sent |
| IPv6 TF | 0 | B | 0 | equal | not-sent |
| IPv6 FL | 0 | B | 0 | equal | not-sent |
| IPv6 L | 0 | B |  | ignore | comp-legth |
| IPv6 NH | 0 | B | 17 | equal | not-sent |
| IPv6 HL | 0 | B | 255 | equal | not-sent |
| IPv6 S Prefix | 0 | U | Alpha::/64 | equal | not-sent |
| IPv6 S IID | 0 | U |  | ignore | DEViid-DID |
| IPv6 D Prefix | 0 | U | Beta::/64 | equal | not-sent |
| IPv6 D Prefix | 0 | U | ::1000 | equal | not-sent |
| UDP S Port | 0 | B | 5683 | equal | not-sent |
| UDP D Port | 0 | B | 5683 | equal | not-sent |
| UDP L | 0 | B |  | ignore | comp-length |
| UDP C | 0 | B |  | ignore | comp-check |

Figure 3. IPv6/UDP rule example

| Rule one | | | | | |
|---|---|---|---|---|---|
| Field ID | Pos | Dir | Target Value | Matching Operator | C/D Function |
| IPv6 V | 0 | B | 6 | equal | not-sent |
| IPv6 TF | 0 | B | 0 | equal | not-sent |
| IPv6 FL | 0 | B | 0 | equal | not-sent |
| IPv6 L | 0 | B |  | ignore | comp-legth |
| IPv6 NH | 0 | B | 17 | Equal | not-sent |
| IPv6 HL | 0 | B | 255 | equal | not-sent |
| IPv6 S Prefix | 0 | U | Alpha::/64 | equal | not-sent |
| IPv6 S IID | 0 | U |  | Ignore | DEViid-DID |
| IPv6 D Prefix | 0 | U | Beta::/64 | equal | not-sent |
| IPv6 D Prefix | 0 | U | ::1000 | equal | not-sent |
| UDP S Port | 0 | B | 5683 | equal | not-sent |
| UDP D Port | 0 | B | 5683 | equal | not-sent |
| UDP L | 0 | B |  | ignore | comp-length |
| UDP C | 0 | B |  | ignore | comp-check |

| Rule two | | | | | |
|---|---|---|---|---|---|
| Field ID | Pos | Dir | Target Value | Matching Operator | C/D Function |
| IPv6 V | 0 | B | 6 | equal | not-sent |
| IPv6 TF | 0 | B | 0 | equal | not-sent |
| IPv6 FL | 0 | B | 0 | equal | not-sent |
| IPv6 L | 0 | B |  | ignore | comp-legth |
| IPv6 NH | 0 | B | 17 | Equal | not-sent |
| IPv6 HL | 0 | B | 255 | equal | not-sent |
| IPv6 S Prefix | 0 | U | Alpha::/64 | equal | not-sent |
| IPv6 S IID | 0 | U |  | Ignore | DEViid-DID |
| IPv6 D Prefix | 0 | U | Beta::/64 | equal | not-sent |
| IPv6 D Prefix | 0 | U | ::1000 | equal | not-sent |
| UDP S Port | 0 | B | 5230 | equal | not-sent |
| UDP D Port | 0 | B | 5230 | equal | not-sent |
| UDP L | 0 | B |  | ignore | comp-length |
| UDP C | 0 | B |  | ignore | comp-check |

**Figure 4. Two rules share the same IPv6 flow in SCHC**

(see Figure 4). This situation appears in all cases that have two or more flows sharing the same header(s). We consider this a memory waste that the constrained devices cannot afford.

As a result, we propose Layered SCHC (LSCHC), a layered context that consists of multiple contexts, rather than the single context for the SCHC. Each context contains rules for a single network layer, with a context for the network layer, a context for the transport layer, and a context for the application layer. The proposed context solution is shown in Figure 5. To identify rules within their respective layer contexts, we divide the rule ID into segments, each segment responsible for identifying the rule used in each context as shown in Figure 5. ALC is the segment for the Application layer context, TLC is the segment for the transport layer context and NLC is the segment for the network context layer. The size of each segment can be set based on the LPWAN technology and the number of rules in each context.

Back to the two flows that are sharing the same IPv6 header example. By splitting the rules between layers, we obtain: a single rule in the NLC, covering the shared IPv6 header; a separate rule for each UDP port, representing the UDP fields of the rules in Figure 4. The rule ID can be constructed as follows. To represent rule one from Figure 4, the LSCHC rule ID should be ALC=0, TLC=1, and NLC =1. To represent rule two from Figure 4, the LSCHC rule ID should be ALC=0, TLC=2, and NLC=1.

LSCHC can save memory on the constrained nodes by storing a single rule for each flow in each layer. Additionally, LSCHC adds flexibility in selecting a suitable rule at the compression side and reduces the processing complexity at the compressor and the decompressor as we will illustrate in the next section.

## 4. Implementation Experiences

We implemented the SCHC scheme in the Contiki-3.0 operating system[1]. Although Contiki targets the IEEE 802.15.4 technology, we argue it provides an adequate tool to test the SCHC. Firstly, because SCHC depends on a static context that does not require synchronization between the network elements, thus, the technology and condition of the channel do not influence its behavior. Furthermore, Contiki provides a good framework to work on and test the SCHC against the currently implemented header compressions in Contiki, which are IPHC for IPv6 and NHC for the next headers. Finally, Contiki already has support for LPWAN technologies[2].

For implementation purposes, we defined a dispatch identifier for the SCHC (3 bits) to be compatible with 6LoWPAN and assumed the size of the rule ID is 5 bits, thus, a device can handle 31 different rules. As recommended by the SCHC draft [8], we used the Concise Binary Object Representation (CBOR) [13] to represent the target value, matching operator, and C/D actions of the SCHC fields. With small modifications, we ported to Contiki the generic CBOR implementation in C that is part of the RIoT [3] operating system for constrained devices. In the following, we provide some insights from our experience in implementing the SCHC and the LSCHC schemes.

### 4.1 Registering Rules

A Device within an LPWAN runs application(s) that produce/accept specific flows. In order to compress/decompress correctly the headers of these flows, the network and the device C/D units must share the same rules and their rule IDs must be identical, otherwise, a mismatch might happen, causing errors in decompressing the packets. However, it cannot be assumed that the network C/D knows in advance all the flows of all applications that can join the network later. Therefore, some initial synchronization is necessary. The draft in [8] does not specify a way for this synchronization to take place. Therefore, we suggest the network C/D supports a layered context with a set of rules that might be suitable for the most common flows. Subsequently, each device within an LPWAN should *offline pick* the most suitable rules for its applications to achieve the maximum compression level. Devices will store a small set of rules they need, and they will use the short rule ID as shown in Figure 5. The network C/D will store a much larger set of rules (covering all applications and more), and therefore will use a larger rule ID. This will require a mapping from device address and device short rule ID to the longer, network C/D, rule ID. Two devices may use the same rule at the network C/D, but with different short rule IDs. It is the short rule IDs that will be sent between a device C/D and the network C/D to specify the used rule in compression.

This proposed method will save memory in the network C/D as well by just saving one version of each rule, rather than building a context specific to each device as proposed in [8]. Furthermore, this method saves bandwidth in the constrained networks by just sending the short identifier among the compression and decompression sides.

---

[1] *Contiki-os.org*.

[2] *github.com/Wi6labs/lorafabian*

[3] *Riot-os.org*.

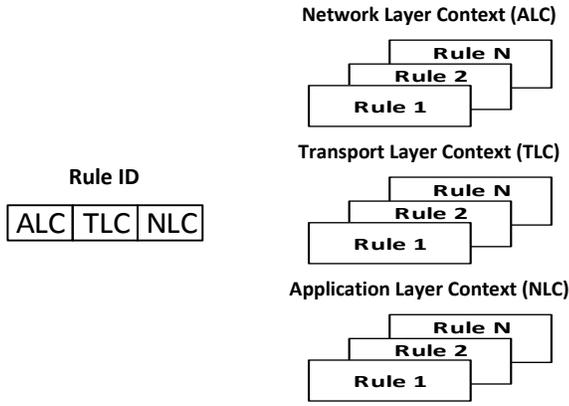

**Figure 5. LSCHC C/D context and its Rule ID**

In the case of a new flow that requires a new rule at the network C/D, a manual administration process is required using an unconstrained connection to the network C/D to register the new rule with a unique long ID. This new rule can then be used by different devices. In all cases, there is no requirements for an *online* learning process between a device C/D and the network C/D, thus saving the limited bandwidth of the LPWANs.

## 4.2 Matching and Selection of Rules

One of the main functions of the C/D unit is to select a suitable rule to compress the header(s). The draft [8] specifies a suitable rule as one where all the rule fields match the packet fields according to the matching operators. If such a rule is found, the packet is processed using the corresponding C/D actions to this rule. Otherwise the packet is sent without compression. However, this is not accurate because the decompressor always expects a rule ID in the packet. Therefore, if there is no matching, the compressor side must send a special ID meaning the packet is not compressed. Furthermore, selecting the first matching rule with the header(s) may not be the best matching approach because there could be more than one rule that matches with the header(s). To get the best solution, the compressor should test all available rules and then select the rule that achieves the best compression factor.

LSCHC, which was presented in subsection 3.1, adds flexibility in selecting the most suitable rule that matches the header(s). Assume a compressor has a rule that targets IPv6/UDP headers as shown in Figure 3. In the case of SCHC, to use this rule, a matching must occur with all fields in the IPv6 header and the UDP header. However, the compressor may produce/receive packets that match with the IPv6 header only or with the UDP header only. With SCHC, the compressor would not be able to use this rule. With LSCHC, because the context is layered, LSCHC can compress the packets that match only with the IPv6 header or match only with the UDP header using the corresponding rule. Therefore, LSCHC is more flexible and can achieve a higher gain in terms of compression factor compared to SCHC in this scenario.

## 4.3 Processing Rules

Processing the context in the case of SCHC is rule specific not field specific because rules target multiple headers, thus, the number of fields in a rule is not identical. We argue that the rule specific processing is not scalable as it requires adding a new processing logic in a case of adding a new rule.

LSCHC solves this problem by separating the layers and processing each layer individually. Isolating the layers makes the processing logic header specific. As all headers in a layer are known e.g. IPv6 in the network layer and UDP, ICMPv6 and TCP in the Transport layer etc. This makes the header specific processing field specific as well. Therefore, processing the context in the case of LSCHC is generic and can work on any kind of rule due to its field specific approach.

## 5. Evaluation and Discussion

In order to evaluate SCHC/LSCHC, we implemented the SCHC/LSCHC scheme in the Contiki-3.0 as mentioned above and emulated it in the Cooja emulator, using the Sky motes, which use Texas Instruments MSP430 microcontroller featuring 16-bit CPU, 8 MHz processor with 10KB of RAM, and 48KB of Flash memory.

The topology setup consists of two Sky motes in which one of them acts as a sender and at the same time it is the root of the Routing Protocol for Lossy Networks (RPL) Destination Oriented Direct Acyclic Graph (DODAG), and the other one acts as a receiver. This one-hop topology is similar to the *star-topology* of LPWANs in which all nodes communicate through a gateway. We used the ipv6/rpl-udp example from Contiki examples in which the sender periodically, every minute, sends "Hello" messages to the receiver. We use small packet sizes (less than 20 bytes) to avoid the effect of packet fragmentation. The ipv6/rpl-udp example produces three different flows between the sender and the receiver. Firstly, an IPv6/ICMPv6 flow with unicast link-local address at the source and multicast with link-local scope address at the destination; this flow is used to discover the neighbors at the beginning of communication. Secondly, an IPv6/ICMPv6 flow with unicast link-local addresses at the source and the destination; this flow is used to create and maintain the RPL DODAG. Thirdly, an IPv6/UDP flow with global addresses at the source and the destination; this flow is used to transmit the "Hello" messages.

To compress the three flows using SCHC/LSCHC, we composed three rules in which each rule targets a flow. For implementation purposes, we defined a dispatch identifier for the SCHC (3 bits) to be compatible with 6LoWPAN and assumed the size of the rule ID is 5 bits, thus, a device can handle 31 different rules and the rule ID that equals 31 is reserved to represent uncompressed packets. We are mainly interested in the compression factor, which is defined as the ratio between the uncompressed size and the compressed size. Higher compression factor indicates higher compression efficiency and a lower amount of data to be sent.

Figure 6 shows the compression factor for each flow in case of SCHC/LSCHC and IPHC/NHC. As shown, SCHC/LSCHC can achieve a higher compression factor than IPHC/NHC in all flows. The first and the second flow are similar (IPv6/ICMPv6). SCHC can compress this flow down to two octets (one octet for dispatch and rule ID and one octet for ICMPv6). As the logic for compressing the ICMPv6 header is not implemented in Contiki, IPHC/NHC compresses these headers down to six octets (two octets for the dispatch and IPHC encoding and four octets for ICMPv6). Regarding the third flow (IPv6/UDP), SCHC/LSCHC can compress the two headers down to one octet (dispatch and rule ID), whereas IPHC/NHC compresses the two headers down to four octets (two octets for dispatch and IPHC encoding, one octet for the stateful compression, and one octet for UDP).

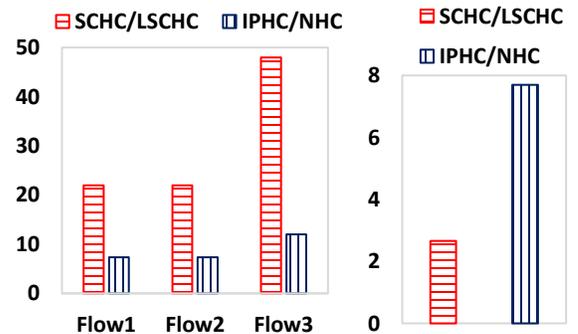

**Figure 6. Compression Factor**    **Figure 7. Octets/packet**

Figure 7 shows the average transmitted octets, the headers only, per packet after running the emulator for six hours. The proportion of the IPv6/UDP and the IPv6/ICMPv6 packets is the same in SCHC/LSCHC (358.33, 301.66) and IPHC/NHC (350, 308.33). We omitted the octets of the MAC frame and the payload from our calculations as they are the same in both header compression schemes. As shown, the SCHC sends on average 2.66 octets/packet for the headers, whereas, IPHC/NHC sends on average 7.69 octets/packet for the headers.

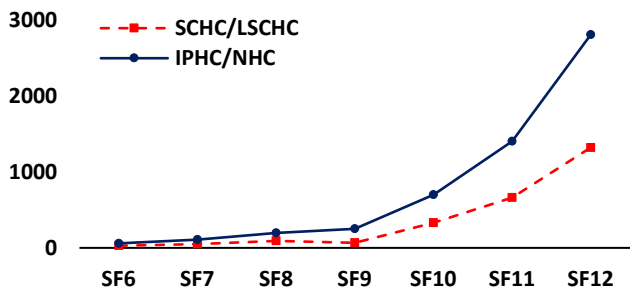

**Figure 8. Total airtime to send the flows header (ms)**

Figure 8 illustrates the total transmission time to send the average flow headers shown in Figure 7 over LoRa [3] for different Spreading Factors (SFs) as an example of a LPWAN technology. Higher SF means more range and better reception, however, also means more transmission time. The calculations are performed for the 125 KHz bandwidth case, 0.10 % duty cycle, 4/5 coding rate and 8 preamble symbols. We used Semtech's LoRa calculator[4], to calculate the transmission times. This metric has direct effect on the power consumption of the devices because the power consumption of low power devices depends very much on the transmission time. Therefore, SCHC/LSCHC helps the devices to live longer with the same power source compared to IPHC/NHC. Also, the results show that in most cases, by using SCHC/LSCHC, the devices can increase the reliability or range of the transmission, while achieving the same transmission time as with IPHC.

## 6. Conclusion

We described SCHC, a new header compression scheme proposed by the IETF to adapt the size of IPv6, UDP and CoAP headers for transmission over LPWANs. Additionally, we presented LSCHC, an enhancement of SCHC that saves memory in constrained devices, reduces the processing complexity and adds flexibility in selecting the matched rule. We implemented and evaluated SCHC/LSCHC compared to IPHC/NHC. The evaluation process showed that SCHC/LSCHC achieves higher compression factors compared to IPHC/NHC and we illustrated the effect of this on the transmission time and the reliability over a LoRa link as an example for LPWAN technology. SCHC/LSCHC is very effective if the potential flows within a network are known in advance, however, it performs poorly with flows that are *unknown* in advance. Therefore, we are currently devising a way to deal with this scenario.

## 7. Acknowledgments

This publication has emanated from research supported by research grants from Science Foundation Ireland (SFI) and is co-funded under the European Regional Development Fund under Grant Numbers 13/RC/2077 and 13/IA/1885.

---

[4] *Semtech.Com/Images/Datasheet/Loradesignguide_STD.Pdf*